\documentclass[%
reprint,
superscriptaddress,
 amsmath,amssymb,
 aps,
]{revtex4-2}

\usepackage{graphicx}
\usepackage{dcolumn}
\usepackage{bm}
\usepackage[mathlines]{lineno}

\begin{document}


\title{Electron round lenses with negative spherical aberration by a tightly focused cylindrically polarized light beam}

\author{Yuuki Uesugi}
    \email{uesugi@tohoku.ac.jp}
    \affiliation{Institute of Multidisciplinary Research for Advanced Materials, Tohoku University, Katahira 2-1-1, Aoba-ku, Sendai 980-8577, Japan}
\author{Yuichi Kozawa}
    \affiliation{Institute of Multidisciplinary Research for Advanced Materials, Tohoku University, Katahira 2-1-1, Aoba-ku, Sendai 980-8577, Japan}
\author{Shunichi Sato}
    \affiliation{Institute of Multidisciplinary Research for Advanced Materials, Tohoku University, Katahira 2-1-1, Aoba-ku, Sendai 980-8577, Japan}

\date{May 30, 2021}

\begin{abstract}
Free electrons moving in an optical standing wave field feel the ponderomotive potential, acting as a refractive-index medium in electron optics.
Emerging technologies involving this potential have been proposed and realized in electron microscopy, such as electron phase-contrast imaging using a laser standing wave in an optical enhancement cavity.
However, the interaction between electrons with a cylindrically distributed optical field has not been investigated although its suitability for electron-optical imaging systems.
In this study, we theoretically show that the divergence and convergence forces are provided by tightly focused cylindrically polarized light beams.
The radially and azimuthally polarized beams with an annular profile are focused using a high numerical aperture optical lens.
The intensity distributions at the focus function are concave and convex electron round lenses, respectively.
The convex lens formed by the azimuthally polarized beam possesses negative (opposite sign) spherical aberration compared with conventional electron round lenses created by electrodes and magnetic coils.
This remarkable result will contribute to the innovative design of electron-optical imaging systems and bring new capabilities into matter-wave optics.

\end{abstract}

\maketitle


When free electrons interact with photons in a laser standing wave, electron scattering tends to be highly directed because one type of photons acts as an incident beam, whereas counter-propagating photons serve as a stimulating beam.
This phenomenon corresponds to stimulating Compton scattering (SCS) in the quantum systems and is known as the Kapitza--Dirac effect (KDE).
Since 1933, when it was first proposed, it has been the subject of theoretical \cite{Kapitza1933,Kibble1966,Fedorov1967,Ezawa1969,Chan1979,Guo1996,Fedorov1997,Li2004,Batelaan2007,Ahrens2013,Kozak2018,Talebi2019} and experimental \cite{Bartell1965,Schwarza1965,Bartell1968,Takeda1968,Bucksbaum1988,Aflatooni2001,Freimund2002,Axelrod2020} research.
Those efforts have brought several electron-optical applications, such as electron phase-contrast imaging \cite{Muller2010,Schwartz2019}, energy modulation and attosecond bunching of electrons \cite{Hilbert2009,Kozak2018-2,Kozak2018-3}, electron bunch-length measurement \cite{Hebeisen2008,Gao2012}, and a proposal of an electron spin-polarizing beam splitter \cite{Dellweg2017}.

The force arising from SCS, as well as ponderomotive forces derived from the second-order motion of an electron in a classical electromagnetic field and electron scattering by an evanescent field, is given by the gradient of a scalar function interpreted as the ponderomotive potential \cite{Fedorov1997}.
This potential acts as a refractive-index medium for electrons, with a shape different from that created by a static electric or magnetic field of which potential distribution is determined by the Laplace equation and boundary conditions on materials, and enables the realization of electron-optical elements with a novel functionality.
Indeed, the use of material-assisted optical fields has led to a wealth of applications in electron microscopy \cite{Barwick2009,Wang2020,Konecna2020,Feist2020} and relevant laser acceleration technology \cite{Wootton2016,Black2019,Schonenberger2019}.

In this study, we focus on the feasibility of a round lens action and spherical aberration correction, which is indispensable for improving the resolution of electron microscopes.
A positive spherical aberration is unavoidable in a round lens using electrodes or magnetic coils, as proven by Scherzer in 1936 \cite{Scherzer1936}.
The convergence force at a distance from the beam axis is stronger than that near the beam axis, resulting in a shorter focal length for peripheral electron trajectories.
Although aberration correctors were developed in the 1990s \cite{Zach1995,Haider1998,Krivanek1999}, they are complicated and difficult-to-use systems consisting of several multipole lenses.
Thus, the development of simpler aberration correction techniques has  attracted significant attention \cite{Konecna2020,Kawasaki2016,Linck2017,Grillo2017}.
Recently, a similar proposal for the aberration collection using a focused paraxial light beam has been investigated \cite{deAbajo2021}. 

We investigate the interaction between electrons with a cylindrically symmetric laser standing wave.
First, we calculate tightly focused cylindrically polarized light beams with an annular profile.
The focused annular beam can create a Bessel-like standing wave at the focal region.
Then, we evaluate force fields derived by the ponderomotive potential.
Furthermore, we calculate the trajectories of the electrons passing through the ponderomotive potential to evaluate the lens characteristics.
This is the first evaluation of the cylindrically symmetric KDE, which is favorable for electron microscopy applications. However, most studies on KDE have postulated light beams lying perpendicular to the electron beam axis.


\begin{figure*}
    \centering\includegraphics[width=17.9cm]{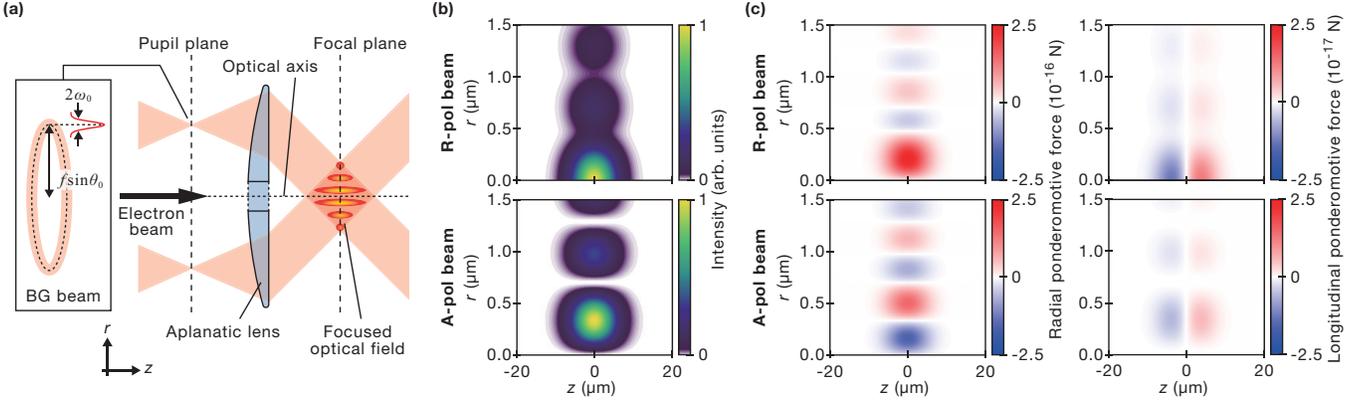}
    \caption{
    \label{fig:fields}
(a) Schematic of the proposed experimental system.
A radially or azimuthally polarized Bessel-Gauss (BG) beam is focused by a high numerical aperture aplanatic lens.
An electron beam propagating along the optical axis passes through the focused field.
The diagram is not drawn to scale.
(b) Intensity distributions of the tightly focused radially (upper) and azimuthally (lower) polarized beams.
These are given by $|u^{\mathrm{R}}_{r}|^{2}+|u^{\mathrm{R}}_{z}|^{2}$ and $|u^{\mathrm{A}}_{\phi}|^{2}$, respectively.
The magnitude is normalized to the maximum value of 1.
(c) Distributions of radial (left) and longitudinal (right) ponderomotive forces.
The directions of the radial forces near the optical axis are reversed in the case of the radially (upper) and azimuthally (lower) polarized beams.
    }
\end{figure*}

The ponderomotive potential and the related force acting on a non-relativistic electron can be expressed as follows:
\begin{gather}
U_{\mathrm{p}}=
    \frac{e^{2}}{4m\omega^{2}}|\mathbf{E}|^{2},\label{eq:up}\\
\mathbf{F}=
    -\mathbf{\nabla}U_{\mathrm{p}}=
    -\frac{e^{2}}{4m\omega^{2}}\mathbf{\nabla}\left|\mathbf{E}\right|^{2},\label{eq:fp}
\end{gather}
where $e$, $m$, $\omega$, and $\mathbf{E}$ are the elementary charge, mass of the electron, angular frequency of light, and electric field of light, respectively.
Equation (\ref{eq:fp}) can be rewritten as
\begin{equation}
    \mathbf{F}
    =-\frac{e^{2}}{4m\omega^{2}}
    \left[
    	    (\mathbf{E}\cdot\mathbf{\nabla})\mathbf{E}^{*}
    	+   \mathbf{E}\times(\nabla\times\mathbf{E}^{*})
    	+   \text{c.c.}
    \right],\label{eq:pondero}
\end{equation}
where c.c. denotes complex conjugate terms.
Here we note the second term on the right side of Eq. (\ref{eq:pondero}), which is the same form as the force arising from SCS in the classical description \cite{Batelaan2007}.
To generate the radial force for the round lens action, either the electric or magnetic field of the light must be oriented in the longitudinal direction.
Such a longitudinal electric (magnetic) field can be created by tightly focusing a radially (azimuthally) polarized light beam with a high numerical aperture (NA) objective lens \cite{Youngworth2000,Zhan2009}.

Light beams represented by any paraxial solution do not possess a longitudinal component; therefore, they cannot be used for the present cases.
The vectorial diffraction theory presented by Richards and Wolf is widely used for accurate descriptions of the tightly focused optical fields \cite{Richards1959}.
We define the electric field of the incident light beam at the pupil of the objective lens that is aplanatic with high NA as
\begin{equation}
    \mathbf{E}_{\mathrm{pup}}(\mathbf{r};t)=E_{0}\mathrm{Re}\left[\mathbf{u}_{\mathrm{pup}}(\mathbf{r})e^{-i\omega t}\right],
\end{equation}
where $E_{0}$ is the peak amplitude of the field, and $\mathbf{u}_{\mathrm{pup}}$ is the normalized vector containing spatial structures of the amplitude, phase, and polarization of the field.
The wavefront over the pupil is transferred to a spherical wavefront by the lens, converging to a focal point on the optical axis.
The field in the focal region can be quantitatively determined by integrating the field expanded into plane waves directing to the focal point within the maximum converging angle, $\alpha$.

With the cylindrical coordinate system, $\mathbf{r}=\mathbf{r}(r,\phi,z)$, we define radially polarized (R-pol) and azimuthally polarized (A-pol) beams as follows:
\begin{align}
    \mathbf{u}_{\mathrm{pup}}^{\mathrm{R}}(r,\phi,z) = \left(p(r)e^{ikz},0,0\right),\\
    \mathbf{u}_{\mathrm{pup}}^{\mathrm{A}}(r,\phi,z) = \left(0,p(r)e^{ikz},0\right),
\end{align}
where $p(r)$ is a real function denoting the amplitude distribution.
It means that both beams are assumed to be propagated along the $z$-axis and have the same profile and a planar wavefront at the pupil.
Applying the vectorial diffraction integration, the tightly focused R-pol beam creates an electric field consisting of radial and  longitudinal components at the focal region, $\mathbf{u}^{\mathrm{R}} = (u^{\mathrm{R}}_{r},0,u^{\mathrm{R}}_{z})$, which are given by
\begin{align}
u^{\mathrm{R}}_{r}&=
    C\int^{\alpha}_{0}d\theta\; P(\theta)\sin2\theta J_{1}(kr\sin\theta)e^{ikz\cos\theta},\label{eq:ur}\\
u^{\mathrm{R}}_{z}&=
    2iC\int^{\alpha}_{0}d\theta\;P(\theta)\sin^{2}\theta J_{0}(kr\sin\theta)e^{ikz\cos\theta},\label{eq:uz}
\end{align}
where $C=\pi f/\lambda$, $f$ is the focal length of the lens, $\lambda$ is the wavelength of light, $J_{n}$ is the Bessel function of the first kind of n-th order, and $P(\theta)$ is the apodization function derived to satisfy the energy conservation between the entrance and exit of the lens.
In the Abbe's sine condition, the apodization function is given by
\begin{align}
    P(\theta)=p(f\sin\theta)\sqrt{\cos\theta}.
\end{align}
Then, the electric field by the A-pol beam is expressed as $\mathbf{u}^{\mathrm{A}} = (0,u^{\mathrm{A}}_{\phi},0)$ and consists of only an azimuthal component given by
\begin{align}
u^{\mathrm{A}}_{\phi}=
    2C\int^{\alpha}_{0}d\theta\;P(\theta)\sin\theta J_{1}(kr\sin\theta)e^{ikz\cos\theta}.\label{eq:uphi}
\end{align}

For calculation, the incident amplitude distribution is defined as an annular ring profile with a Gaussian-shaped width by
\begin{equation}
    p(r) = \exp\left[-\frac{(r-f\sin\theta_{0})^{2}}{w_{0}^{2}}\right],
\end{equation}
where $\theta_{0}$ is the cone angle (half angle) of beam focusing, and $w_{0}$ is the half width of the annular ring.
Such a beam produces a Bessel beam with a finite extension in the focal region, known as a Bessel-Gauss (BG) beam.
It is worth noting that there are analytic solutions for paraxial BG beams, and non-paraxial BG beams are represented by a more complicated series solutions \cite{Gori1987,Borghi2001,Gawhary2010}.
The BG beam coincides with the Bessel beam in the limit where the annular ring size is infinitesimal.
As the cone angle approaches $\pi/2$, the period of the transverse oscillation approaches about half the wavelength.
Figure \ref{fig:fields}(a) shows a schematic of the geometrical configuration of our optical system.
A small $w_{0}$ increases the interaction length with electrons, and a large $\theta_{0}$ causes efficient KDE.


Figure \ref{fig:fields}(b) shows the calculated normalized intensity distributions: $|\mathbf{u}^{\mathrm{R}}|^{2}=|u^{\mathrm{R}}_{r}|^{2}+|u^{\mathrm{R}}_{z}|^{2}$ and $|\mathbf{u}^{\mathrm{A}}|^{2}=|u^{\mathrm{A}}_{\phi}|^{2}$ for the focused R- and A-pol beams, respectively.
Table \ref{tab:fields} presents the parameters used for calculations.
The intensity of the R-pol beam has a maximum magnitude on the optical axis because the longitudinal component depends on the 0-th order Bessel function and is dominant.
However, the intensity of the A-pol beam has a null on the axis.
Figure \ref{fig:fields}(c) shows the calculated radial and longitudinal components of the forces derived from the gradient of the ponderomotive potential, $U_{\mathrm{p}}^{\mathrm{R,A}}=e^{2}E_{0}^{2}/4m\omega^{2}|\mathbf{u}^{\mathrm{R,A}}|^{2}$.
The optical power of the incident beam is set to 100 W in both cases.
The focused R-pol beam produces a radial divergence force, while the focused A-pol beam produces a convergence force on electrons around the optical axis.
The magnitude of the radial force is $\sim20$ times greater than that of the  longitudinal component.
This is because bumps of the intensity in the transverse direction have an interval of about half the wavelength, while in the longitudinal direction, the intensity spreads at a distance by more than ten times compared with the wavelength.

\begin{figure*}
    \centering\includegraphics[width=17.9cm]{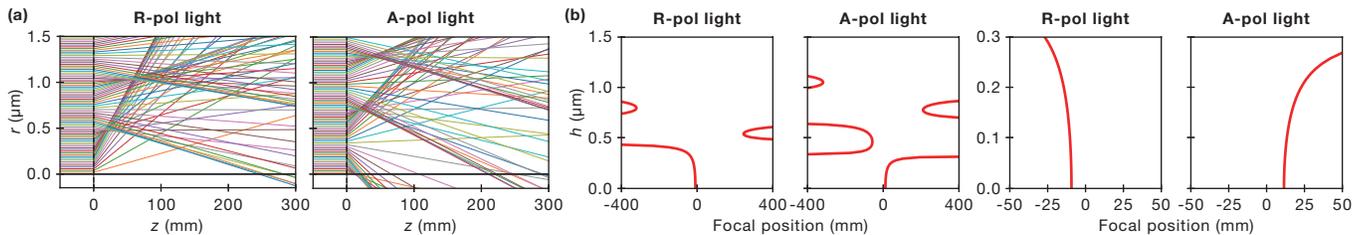}
    \caption{
    \label{fig:rays}
(a) Trajectories of electrons entering the force field derived by the divergent of the ponderomotive force parallel to the optical axis.
The field is distributed around $z=0$.
Near the optical axis, a concave lens action occurs in the case of the radially polarized beam (left), and convex lens action occurs in the case of the azimuthally polarized beam (right).
(b) Focal positions of each electron trajectories in both the radially and azimuthally polarized light cases.
The vertical axis is the incident electron distance from the optical axis.
The set of two plots on the right is an enlarged version of the set on the left.
    }
\end{figure*}

To evaluate the characteristics of these \textit{ponderomotive lenses}, we calculate trajectories of electrons by solving the equation of motion with the obtained force field, corresponding to the ray-tracing analysis in optics.
The electrons are injected into the field parallel to the optical axis at different distances from the axis.
The electron energy is set to 1 keV, and the electron velocity is assumed to be constant in the field.
Figure \ref{fig:rays}(a) shows the results of the trajectory calculations.
The central part of the ponderomotive lens acts as concave and convex round lenses when focusing the R- and A-pol beams, respectively.
The fringes of the intensity distribution form annular ring lenses.

Figure \ref{fig:rays}(b) shows the focal lengths and the spherical aberration characteristics of the obtained ponderomotive lenses.
The $z$-position where the electron trajectory intersects the optical axis is plotted as a function of the incident electron distance from the optical axis, $h$.
The most remarkable result we obtained is the negative (opposite sign) spherical aberration of the convex round lens generated by the focused A-pol beam; the convergence force for peripheral rays is weaker than that for paraxial rays.
This characteristic cannot be obtained with conventional round lenses based on cylindrically symmetric electrodes and magnetic coils.
The curve, which diverges infinitely at $h\sim0.3$ $\mathrm{\mu m}$, suggests that the lens can be adapted for correcting the third- and higher-order spherical aberrations.
Although the negative focal length created by the focused R-pol beam is a unique property not found in the conventional round lenses, it may be unlikely to be applied to electron-optical imaging systems due to the positive (normal sign) spherical aberration.

\begingroup
\squeezetable
\begin{table}
    \centering
    \caption{
Parameters used in the calculation of the tightly focused cylindrically polarized light beams.
    }
    \label{tab:fields}
    \begin{tabular}{cccc}
\hline\hline
Parameter & Symbol & Value & Unit\\
\hline
Wavelength & $\lambda$ & 1 & $\mathrm{\mu m}$\\
Ring half width & $w_{0}$ & 50 & $\mathrm{\mu m}$\\
Cone angle & $\theta_{0}$ & 60 & degrees\\
Focal length & $f$ & 2 & mm\\
Max. converging angle & $\alpha$ & $\theta_{0}\times1.1$ & degrees\\
Optical power & - & 100 & W\\
Electron energy & - & 1 & keV\\
\hline\hline
    \end{tabular}
\end{table}
\endgroup

\begin{figure*}
    \centering\includegraphics[width=17.9cm]{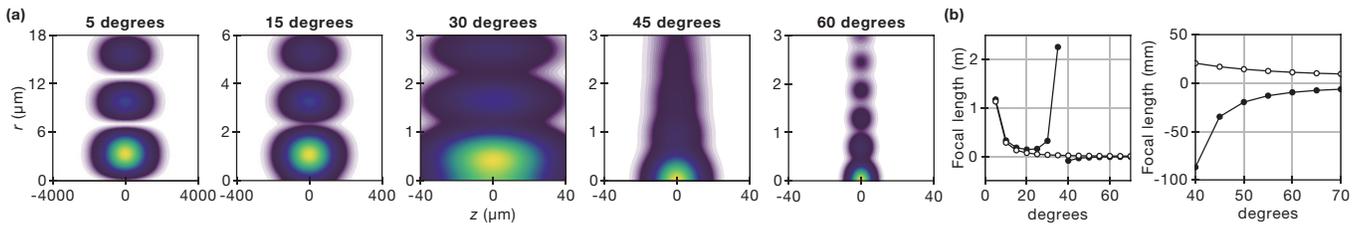}
    \caption{
    \label{fig:nas}
(a) Intensity distributions of the focused radially polarized beam for several values of the cone angle of beam focusing.
They are obtained using the parameters in Table \ref{tab:fields}, except for $theta_{0}$.
As the cone angle decreases, the  longitudinal component $u^{\mathrm{R}}_{z}$ that forms intensity on the optical axis becomes weaker.
(b) Paraxial focal length of the ponderomotive lens as a function of the cone angle.
The filled circles are for the radially polarized beam, while the open circles are for the azimuthally polarized beam.
The solid lines are guides to the eye.
    }
\end{figure*}

We can see that the pupil diameter of the ponderomotive round lens is $\sim\lambda$, indicating that it can be increased using a longer wavelength of light.
A large pupil diameter is favorable for practical applications in electron microscopy because it facilitates the design and adjustment of an electron-optical imaging system combined with a circular aperture stop.
A smaller aperture with diameters of less than one micrometer tends to cause hole geometry instability due to contamination and thermal drift.
However, we obtained that longer wavelengths increase the focal length due to the decrease in the optical power density.
The focal length is also proportional to the input optical power according to Eq. (\ref{eq:up}) and the velocity of the incident electrons.
Eventually, the NA of the ponderomotive lens depends mainly on the input optical power and the electron energy.

Next, we examine the dependence of the round lens action on the cone angle of BG beam focusing.
Figure \ref{fig:nas}(a) shows the intensity distribution of the focused R-pol beam calculated by changing $\theta_{0}$.
As the cone angle decreases, the transverse oscillation period increases and the magnitude of the longitudinal component decreases.
Figure \ref{fig:nas}(b) shows the plots of the paraxial focal length of the round lens action.
The filled circles are for the focused R-pol beam, while the open circles are for the focused A-pol beam.
In the former, the sign of the focal length discontinuously reverses when $\theta_{0}$ is between 35 and 40 degrees.
The discontinuous point, where the lens power extinguishes, may indicate an equilibrium between the force arising from SCS and the ponderomotive force.
For the convex lens action, the focal length increases monotonically with decreasing $\theta_{0}$, and the change is relatively gradual.

The results showed that a tightly focused radially (azimuthally) polarized BG beam creates cylindrically symmetric ponderomotive potential and acts as a concave (convex) electron round lens with $\sim\pm10$ mm focal length.
The negative spherical aberration generated by the convex lens will bring a new type of spherical aberration compensation in electron microscopy.
In particular, it will be effective for scanning electron microscopy (SEM) with low-energy electron beams.
As shown in Fig. \ref{fig:rays}(b), we roughly estimate that the third-order spherical aberration can be eliminated by using an appropriate aperture stop, and the convergence semi-angle of more than 10 mrad can be obtained in a typical SEM system.

Starting from a linearly polarized Gaussian beam, the techniques of shaping the beam into an annular profile using a spatial light modulator and converting it into R-/A-pol light using some special waveplate are very common \cite{Kozawa2019,Kozawa2020}.
The optical power of 100 W employed in the calculation was chosen as a practical limit considering the damage threshold of each optical element for a continuous-wave laser output.
However, a higher optical power will be available using a pulsed laser, e.g., picosecond pulses with a peak power of 1 MW are readily accessible.
Therefore, the lens action for $\sim100$ keV electron beams of a medium voltage electron microscope would be feasible, even if the relativistic corrections are considered \cite{Axelrod2020}.
The use of such short-pulsed lasers is well adapted for applications in ultrafast electron microscopy \cite{Barwick2008,Piazza2013,Kieft2015,Feist2017,Houdellier2018,Arbouet2018}.
Notably, KDE also occurs on beams of atoms and molecules \cite{Salomon1987,Kunze1996,Nairz2001,Eilzer2014}.
Thus, the use of the ponderomotive lens in matter-wave optics and the focused ion beam technology is also promising.


We thank Dr. T. Kawasaki for the discussion on this work.
This work was supported by JSPS KAKENHI Grant Number JP20H02629 and Research Foundation for Opto-Science and Technology.


\bibliography{apssamp}

\end{document}